%% file: main.tex
\documentclass[sigconf]{acmart}
\usepackage[linesnumbered,ruled,vlined]{algorithm2e}
\AtBeginDocument{%
  }

\copyrightyear{2026}
\acmYear{2026}
\setcopyright{cc}
\setcctype{by}
\acmConference[ICCAD '26]{IEEE/ACM International Conference on Computer-Aided Design}{November 08--12, 2026}{San Jose, CA, USA}
\acmBooktitle{IEEE/ACM International Conference on Computer-Aided Design (ICCAD '26), November 08--12, 2026, San Jose, CA, USA}
\acmDOI{10.1145/3831252.3834110}
\acmISBN{979-8-4007-2873-0/2026/11}

\begin{document}

\title{L\textsuperscript{2}C\textsuperscript{2}: Enhancing \underline{L}arge-scale \underline{C}hiplets for Agent \underline{L}ong-\underline{C}ontext LLM Inference with Sparse and Distributed Attention}

\author{Kanbo Bai}
\email{2301111810@stu.pku.edu.cn}
\affiliation{
   \institution{Peking University}
   \city{Beijing}
   \country{China}}

\author{Zhonghua Mu}
\email{2300012807@stu.pku.edu.cn}
\affiliation{
   \institution{Peking University}
   \city{Beijing}
   \country{China}}

\author{Hongze Zhou}
\email{2401210560@stu.pku.edu.cn}
\affiliation{
   \institution{Peking University}
   \city{Beijing}
   \country{China}}

\author{Tianyu Jia}
\authornote{Corresponding author.}
\email{tianyuj@pku.edu.cn}
\affiliation{
   \institution{Peking University}
   \city{Beijing}
   \country{China}}

\begin{abstract}
Long-context inference is a rising demand for LLMs, especially for agent applications. Sparse attention (SpAt) and chiplet technology are promising solutions to reduce computations and scale hardware performance. However, deploying SpAt on distributed chiplets necessitates context parallelism for workload partitioning, introducing complex distributed attention. 
In this work, we present L\textsuperscript{2}C\textsuperscript{2}, a system optimization for distributed SpAt in long-context inference. During the partial-prefill stage, we propose a novel Split-Comm Attention mechanism that leverages the high-bandwidth interconnect of chiplets to trade communication overhead for reduced memory access. We combine it with conventional Ring Attention to form a Bi-Mode method that adapts to varying input sequence lengths. During the decoding stage, we analyze the cause of memory access imbalance and address it by static interleaved token assignment. Dynamic agile cache balancing is also proposed to adapt to SpAt skewness. Evaluations show our optimization can achieve 1.45$\times$ and 1.53$\times$ speedup in partial-prefill and decoding with SpAt on large-scale chiplets.
\end{abstract}



\maketitle

\input{1_introduction}

\input{2_background}

\input{3_motivation}
\input{4_design}
\input{5_evaluation}

\begin{acks}
This work was supported in part by NSFC Grant 92464202 and Beijing NSF Grant L2603003.
\end{acks}

\bibliographystyle{ACM-Reference-Format}
\bibliography{refs}
\end{document}

%% file: 1_introduction.tex
\section{Introduction}
\begin{figure}[t]
  \centering
  \includegraphics[width=\linewidth]{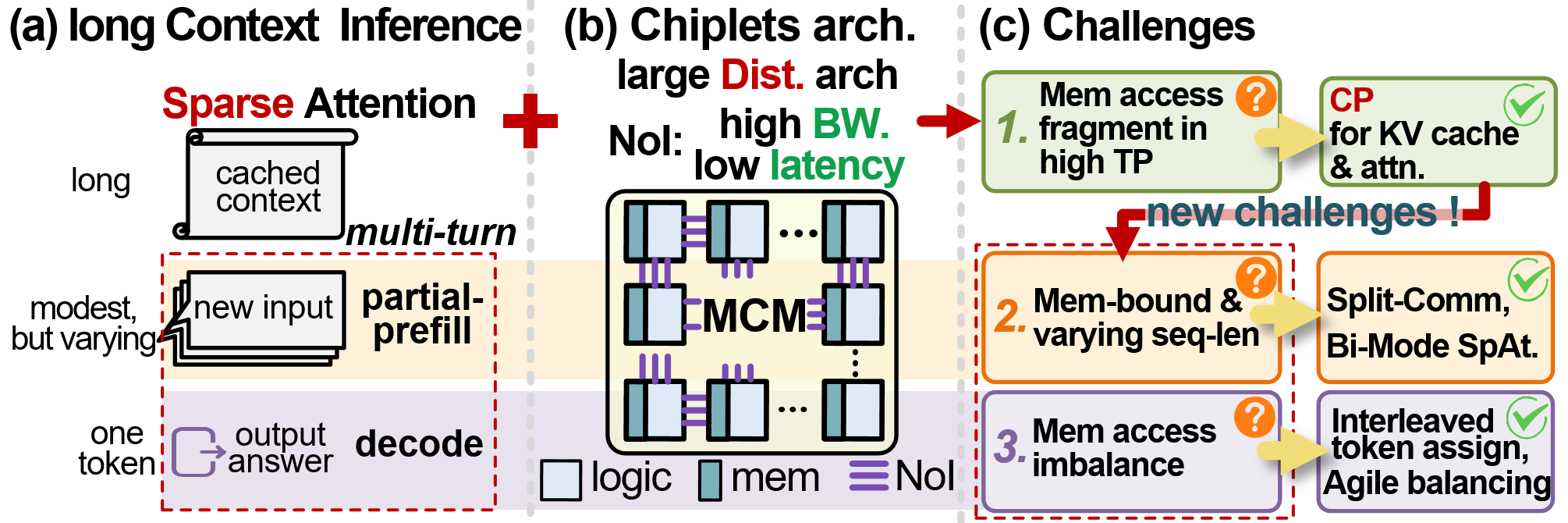}
  \vspace{-15 pt}
  \caption{Challenges of long context LLM on chiplets.}
  \label{fig:intro}
  \vspace{-15 pt}
\end{figure}

Long-context window substantially improves the capability of large language models (LLMs), and the rising demand for AI agent \cite{agent} applications makes long-context inference increasingly important.
At the software level, long-context inference is dominated by attention, which exhibits two emerging features.
First, \textit{sparse attention} (SpAt) is prevalent to mitigate computation overhead by pruning the attention to focus on critical token pairs, such as DeepSeek NSA \cite{nsa}. Second, agent applications drive long context inference into multi-turn workflows as shown in Fig.~\ref{fig:intro}(a). Based on cached context, each turn consists of a \textit{partial-prefill} stage for an input sequence and an iterative \textit{decoding} stage for reasoning. Generally, the context is long and accumulates over turns. Input sequences are typically modest in length, but a few long sequences also introduce variability that must be accounted for. 

At the hardware level, chiplet has emerged as a promising technology to scale up the hardware performance for larger AI workloads. 
As shown in Fig.~\ref{fig:intro}(b), by integrating numerous chiplets onto a package substrate, a multi-chiplet-module (MCM) is established. The area of each chiplet can be reduced, improving manufacturing yield. But the overall integration scale of MCM can be extended, even up to wafer scale \cite{dojo,WSC_LLM}. And the network-on-interposer (NoI) interconnects all chiplets in a high-bandwidth and low-latency communication domain. In conventional GPU clusters, a node has 8 GPUs, with intra-node and inter-node communication is limited by 900 GB/s NVLink \cite{nvidia2022hopper} and 50GB/s InfiniBand (IB) links \cite{deepseekai2025deepseekv3technicalreport}. In MCMs, all chiplets can benefit from a NoI with multi-TB/s bandwidth \cite{dojo} and ns-scale latency \cite{NoI_lc}. 
However, the highly distributed architecture of large-scale MCM complicates the deployment of SpAt-enabled long-text LLMs.
Prior work \cite{h2eal} for distributed memory employs high-degree tensor parallelism (H-TP), which partitions attention and KV caches along the head dimension. 
But this approach incurs severe fragmented memory access for KV caches in SpAt.

Among different parallelisms, context parallelism (CP) partitions the token storage and attention computation along the sequence length and is 
more suitable than H-TP for the deployment of SpAt in MCM architecture. But CP introduces \textit{distributed attention}, which is typically implemented by Ring Attention \cite{ring_attention} mechanism, with new challenges in different stages, as shown in Fig.~\ref{fig:intro}(c). 
For partial-prefill, the runtime dynamism of context and input length causes shifting latencies across computation, memory access, and communication. 
In particular, SpAt reduces compute intensity, thereby rendering memory access more likely to be a limiting factor.
For decoding, although it always processes only a single token per iteration, SpAt induces significant memory access imbalances across the spatially distributed chiplets, decreasing total efficiency.

How to leverage the communication advantages of MCM to address memory-related challenges in partial-prefill and decoding is underexplored.
In this work, we analyze the deployment complexity of long-context inference on large-scale MCM and propose system optimizations. For partial-prefill, we observe that typical short-input with long-context cases in agent applications are memory-bound, whereas existing Ring Attention mechanisms may incur memory access redundancy. Unlike Ring Attention, which overlaps communication with attention processing, we separate them and propose the \textit{Split-Comm} mechanism to reduce memory access. This trades non-overlapped communication for memory efficiency, leveraging the high-bandwidth NoI in MCM. We employ a Bi-Mode method combining Split-Comm and Ring Attention to handle the dynamics of partial-prefill, as Split-Comm yields no gains for rare long-input cases. 
For decoding, we reveal the inherent cause of the memory access imbalance and adopt a static interleaved token assignment to mitigate it based on the features of SpAt. To further minimize imbalances stemming from dynamic SpAt patterns, we propose the \textit{agile cache balancing} that rapidly adapts to shifting SpAt focus with latency-hidden KV cache migration, leveraging the low latency advantage of NoI. Evaluations on ASTRA-sim simulator show our solution offers 1.53$\times$ and 1.45$\times$ speed up for decoding and partial-prefill for long context inference on MCM.
Although our work targets large-scale chiplets, it can be extended to other distributed architectures, e.g., GPU clusters. It is particularly relevant to cutting-edge supernode clusters scaling toward larger and higher-bandwidth domains, such as NVL72 \cite{nvidia2024blackwell}.

The contributions of this work are summarized as follows:
\begin{itemize}
\item We identify memory fragmentation as a major issue of H-TP in SpAt, introducing CP in long-context inference under highly distributed MCM.
\item In the partial-prefill stage, leveraging the high-bandwidth NoI, we propose Split-Comm attention, which trades off minimal communication overhead for significant reductions in memory access. We combine it with Ring Attention to accommodate varying input sequence lengths during partial-prefill.
\item In the decoding stage,  we employ static interleaved token assignment to resolve the memory access imbalance, complemented by dynamic agile cache balancing leveraging the low-latency NoI interface to adapt to shifting SpAt focus.
\end{itemize}

%% file: 2_background.tex
\section{Background}
\begin{figure}[t]
  \centering
  \includegraphics[width=0.95\linewidth]{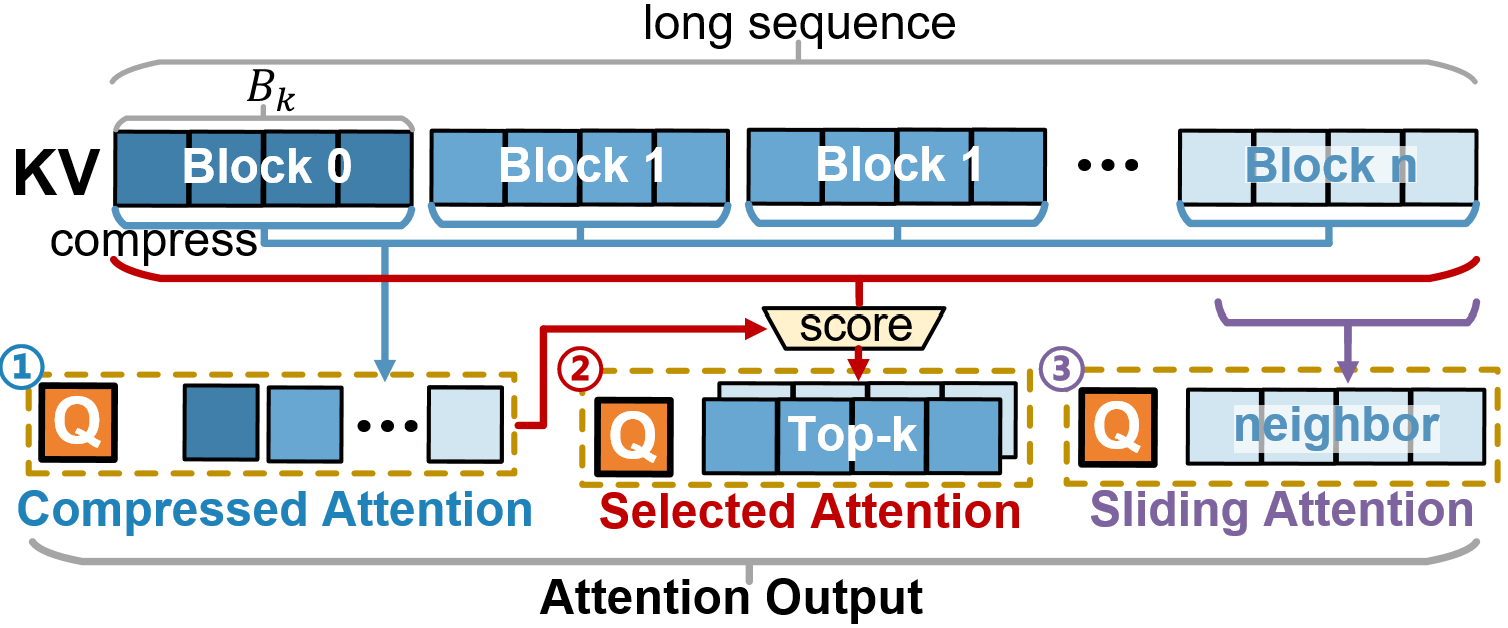}
  \vspace{-10 pt}
  \caption{Mechanisms of native sparse attention (NSA).}
  \label{fig:nsa}
  \vspace{-15 pt}
\end{figure}
\subsection{Sparse Attention}
In full attention, the query (Q) of each token attends to the keys (K) and values (V) of all preceding tokens. 
\textit{Sparse attention} optimizes the computation by allowing each Q to attend to only a subset of KVs. 
Our work is based on DeepSeek-NSA \cite{nsa}, which is shown in Fig.~\ref{fig:nsa}. The KV of a long sequence is divided into multiple KV blocks, each containing $B_{k}$ tokens. NSA is composed of three attention mechanisms:
(1) Compressed attention, where each KV block is compressed into a single token, and each Q attends to all compressed KV in the preceding context. 
(2) Selected attention, where each Q attends to the top-$k$ KV blocks based on the attention scores from the compressed attention. The value of $k$ is chosen based on the sequence length and sparse ratio. (3) Sliding attention, where Q attends to the KVs within a neighbor sliding window. The attention output is the summation of the three mechanisms. 

The three mechanisms in NSA serve distinct purposes. Compression captures coarse-grained global information, whereas sliding focuses on neighboring context. Selection achieves dynamic SpAt to extract the critical information. It is the core of NSA and accounts for the majority of computation and latency \cite{fsa}. For example, with a 128k context length  ($B_k=32$, sparse ratio=0.9), the selection invloves 3 $\times$ and 25 $\times$ more tokens than the compressed and sliding attention. Variations in mechanism and granularity of selection among other SpAt do not affect our following analysis.


\subsection{Parallelism and Distributed Attention}
\label{sec:para}
\begin{figure}[t]
  \centering
  \includegraphics[width=0.95\linewidth]{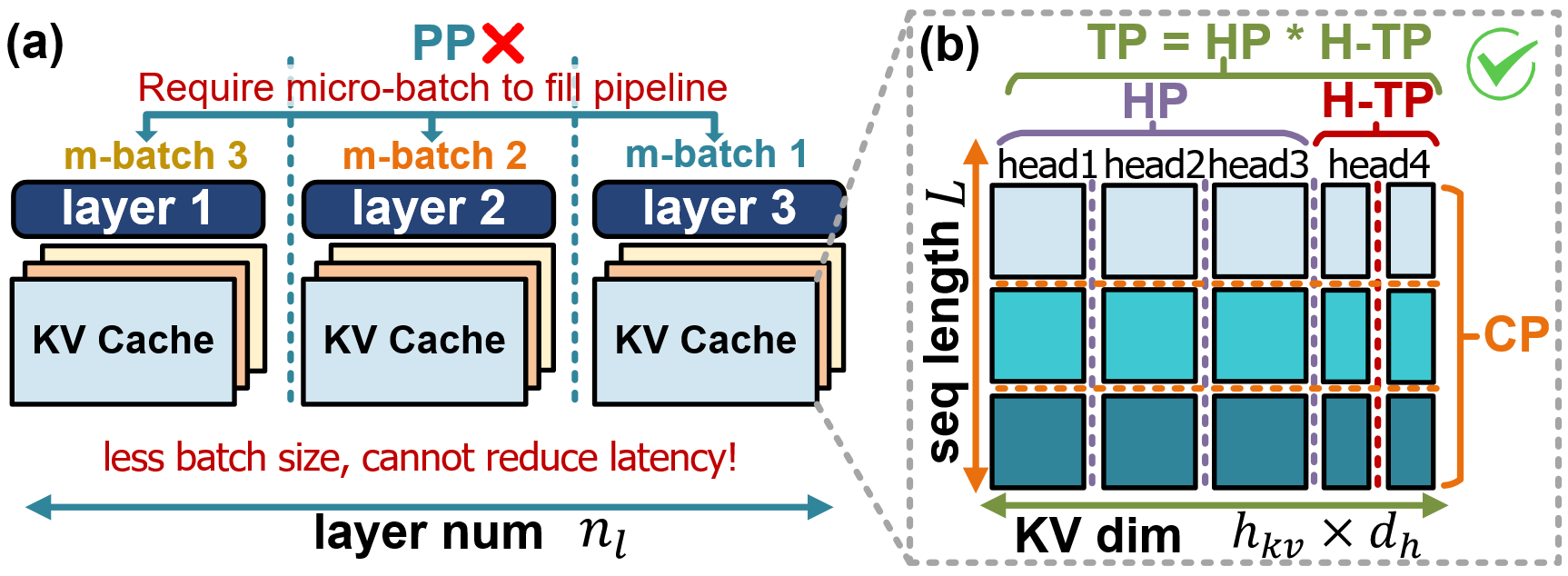}
  \vspace{-10 pt}
  \caption{Introduction of diverse parallelism strategies. } 

  \label{fig:para}
  \vspace{-5 pt}
\end{figure}

The distribution of computation and KV cache across different chiplet dies relies on diverse parallelisms. 
We use the KV cache as an example to list the splittable dimensions:
\begin{equation}
\label{cp-ep}
KV\_size = 2 \times L \times n_{l} \times h_{kv} \times d_{h}
\end{equation}
where $L$, $n_{l}$, $h_{kv}$, $d_{h}$ denote the sequence length, number of layers, number of KV heads, and head dimension, respectively. Different parallelisms shard workload along these dimensions.

\begin{figure}[t]
  \centering
  \includegraphics[width=\linewidth]{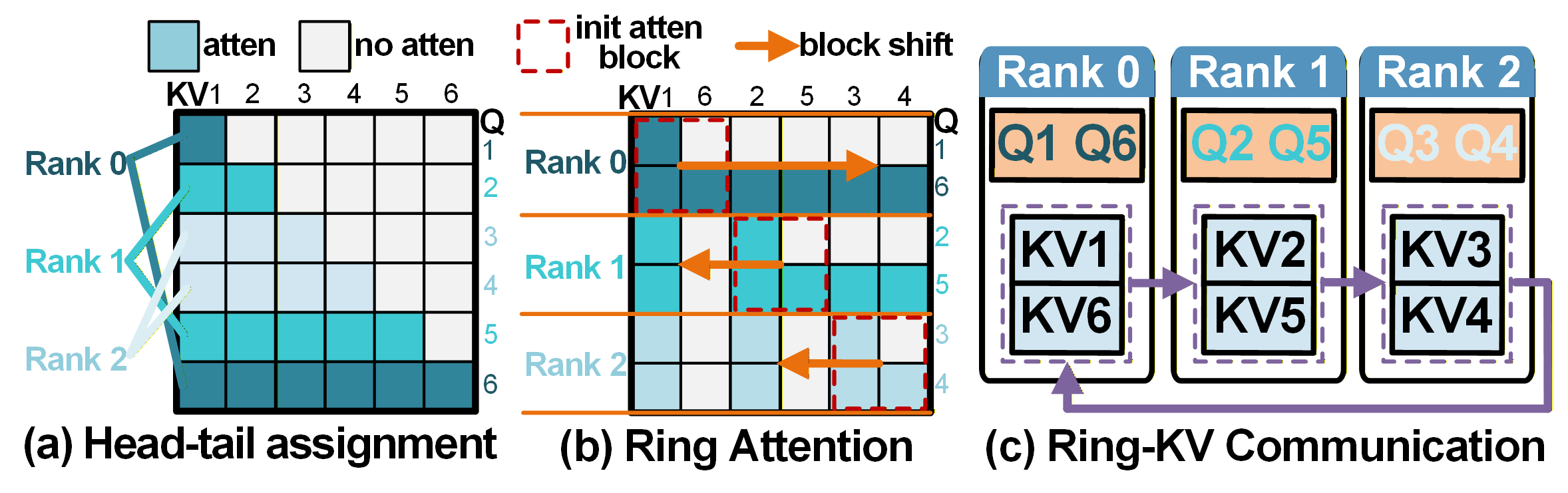}
  \vspace{-15 pt}
  \caption{Context parallelism and Ring Attention scheme. 
  }
  \label{fig:cp}
  \vspace{-15 pt}
\end{figure}

As shown in  Fig.~\ref{fig:para}(a), the pipeline parallelism (PP) performs model partitioning along layer number $n_{l}$. However, PP only increases inference throughput but does not reduce latency, and it requires splitting the batch into micro-batches to fill the pipeline, which reduces the batch size and causes redundant weight access. Therefore, PP is less commonly used in LLM inference serving \cite{WEAVER}. 


As shown in  Fig.~\ref{fig:para}(b), tensor parallelism (TP) splits computation and memory along the KV heads $h_{kv}$, also known as head parallelism (HP). But common GQA \cite{gqa} LLMs have a limited $h_{kv}$, e.g., $h_{kv}=8$ in Llama 3 \cite{llama3}. Therefore, high-degree TP (H-TP) further splits along the head dimension $d_{h}$
\cite{h2eal}.
Context parallelism (CP), which is originally designed for training and then extended to inference \cite{cp}, partitions the token storage and attention along the sequence length $L$.  

As shown in Fig~\ref{fig:cp}(a), CP introduces \textit{distributed attention}, where
each colored block denotes the attention between a Q (column) and a KV (row), with different colors marking the CP rank (i.e., dies in MCM) responsible for the computation. Here we consider only the full-attention case and do not apply SpAt. It can be seen that each rank is assigned a subset of tokens, computing the attention using its Q against all KV. 
Under causal masking, later tokens attend to more preceding tokens. 
The widely used \textit{head-tail} token assignment \cite{loongTrain} packs tokens from the beginning and end of the sequence into the same rank to ensure workload balance. 
Based on online-softmax \cite{online}, attention can be decomposed into attention blocks. 
As shown in Fig~\ref{fig:cp}(b), tokens can be initially distributed across ranks with head-tail assignment, and each CP rank computes the attention block via the \textit{Ring Attention} \cite{ring_attention} shown in the red box.
In the next timestep, these attention blocks are shifted along the orange arrows. In parallel, ranks keep their own Q and exchange the stored KV via ring communication as shown in Fig~\ref{fig:cp}(c). This timestep is repeated for CP iterations to complete the Ring Attention.

%% file: 3_motivation.tex
\section{Architecture and Analysis}

\subsection{Chiplets and Parallelism Strategies}

\begin{figure}[t]
  \centering
  \includegraphics[width=\linewidth]{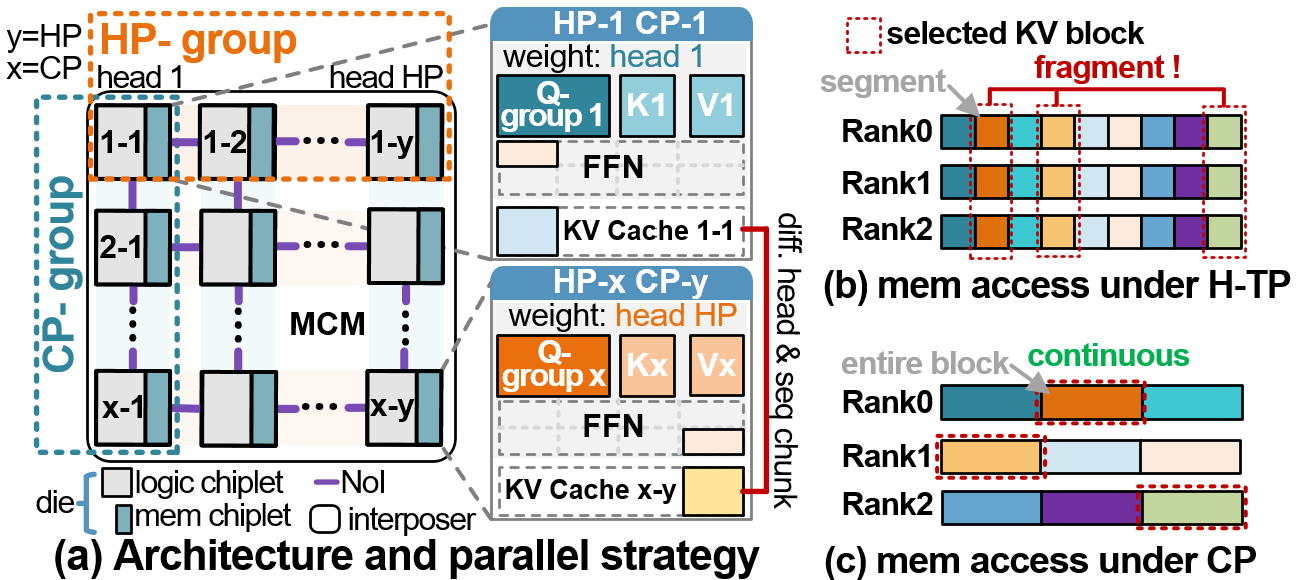}
  \vspace{-15 pt}
  \caption{MCM architecture and parallel strategies. } 
  \label{fig:arch}
  \vspace{-15 pt}
\end{figure}

Fig.~\ref{fig:arch}(a) shows the chiplet architecture adopted in this work, based on prior large-scale MCM designs. \cite{WSC_LLM,memplex}. A logic chiplet and memory chiplets are coupled together for local data access. Following \cite{WSC_LLM}, we refer to such a couple as a die, even though it actually comprises multiple chiplets. Each logic chiplet is typically smaller than the reticle limit to improve yield, so MCM requires more dies than a GPU cluster to achieve comparable scaling. For example in \cite{WSC_LLM}, a logic chiplet has an area of 500$\mathrm{mm}^2$, and up to 63 such chiplets are integrated, each equipped with 2 HBM2e memory chiplets. Dies are interconnected via a network-on-interposer (NoI), which typically uses a mesh topology, constituting an ($x,y$) die array. We assume that the bandwidths of the NoI and the memory chiplets are $BW_{c}$ and $BW_{m}$, and the compute performance of the logic chiplet is $C$.

\label{arch}

To deploy LLM inference on large-scale chiplets, the primary consideration is choosing an appropriate parallel strategy.
As mentioned in Sec.~\ref{sec:para}, TP (including HP and H-TP) and CP are two options. The parallelism needs to match the count of dies:
\begin{equation}
\label{TP_CP}
x \times y = TP \times CP = HP \times H\text{-}TP \times CP
\end{equation}

Since attention across heads is independent, HP is always the preferred parallelism. However, in common GQA LLM, the limited head counts conflict with the large number of dies in MCMs, forcing us to adopt H-TP or CP to further increase the parallel degree. 

CP has clear advantages over H-TP in SpAt for two reasons. First, H-TP introduces a large communication overhead. Prior work \cite{cp} indicates that introducing CP can reduce communication volume compared to H-TP. Second, H-TP can cause severe memory access fragmentation in SpAt 
As shown in the Fig.~\ref{fig:arch}(b), under H-TP, each memory die stores a small segment for every KV block
since its KV dimension is only $h_{kv}/H\text{-}TP$. In SpAt, especially during decoding with a single input token, only a small subset of non-contiguous KV blocks are selected. DRAM performs poorly when repeatedly reading such small, scattered pieces of data.
As shown in Fig.~\ref{fig:arch}(c), under CP, each rank stores a subset of KV blocks, while every block is kept entirely. This ensures contiguous access to an entire KV block, improving memory efficiency.

Fig.~\ref{fig:arch}(a) also illustrates the implementation of our parallel strategy. For simplicity, we adopt CP and HP parallelism groups to the rows and columns of the die array, respectively.
The ($i, j$)-th die stores the KV cache of the $j$-th head in the $i$-th slice of the sequence. The weights of QKV generation are replicated across CP ranks, so the ($i, j$)-th die holds the weights of the $j$-th KV head and $j$-th Q head group. The other weights, e.g., FFN, are independent of the attention heads and are partitioned by rows and columns across all dies. The right side of Fig.~\ref{fig:arch}(a) shows the stored weights and KV cache of two example chiplet dies.

\subsection{Analysis of Distributed Sparse Attention}
\label{sec:challenge}
\begin{figure}[t]
  \centering
  \includegraphics[width=\linewidth]{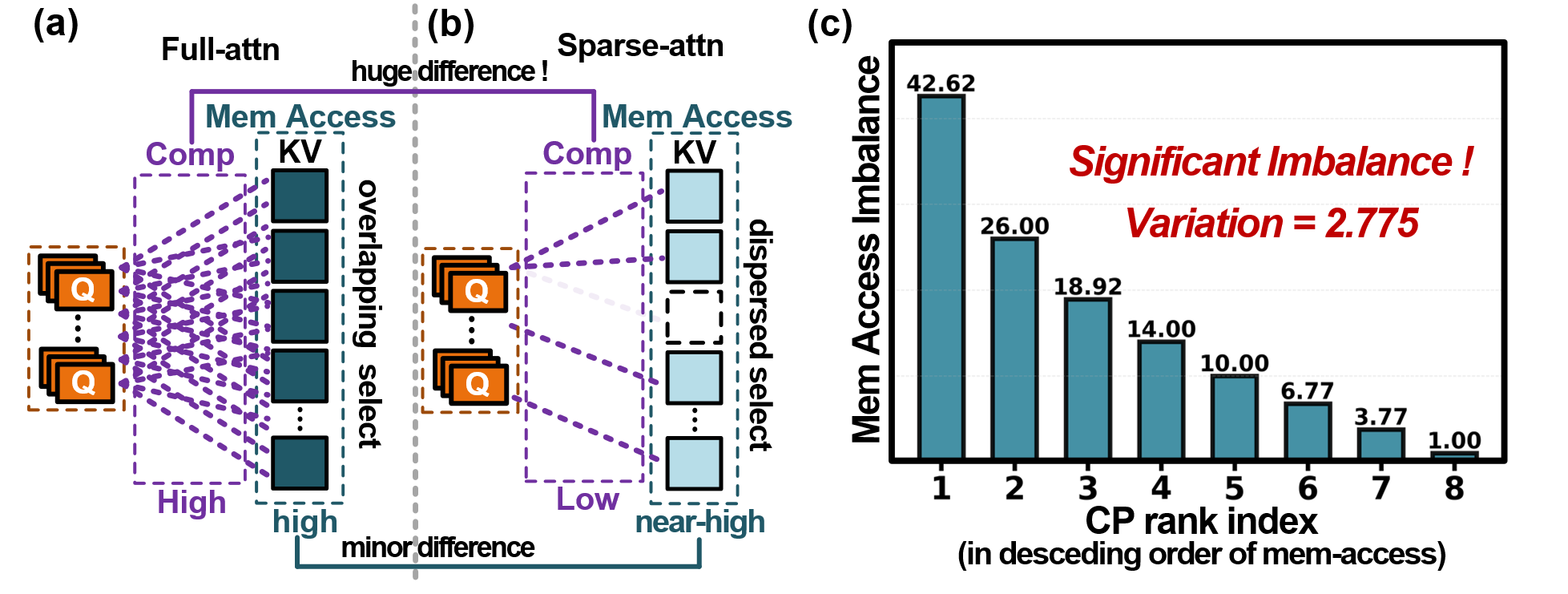}
  \vspace{-15 pt}
  \caption{Challenges of distributed SpAt on MCM.}
  \label{fig:challenge}
  \vspace{-10 pt}
\end{figure}
Although CP alleviates the KV cache fragmentation, it also introduces distributed attention as tokens span multiple ranks. Ring Attention in Sec.~\ref{sec:para} is designed for full attention without sparsity, while the SpAt complicates the distributed attention scenarios. In multi-turn long-text inference, partial-prefill and decoding account for the majority of latency due to extensive repetitions; however, each of them faces distinct challenges under distributed SpAt. 

For partial-prefill of each turn, SpAt makes memory access more likely to become the bottleneck. Fig.~\ref{fig:challenge}(a) shows the reason. In full attention, each Q attends to all KV, and the numerous purple links illustrate the heavy computation. In SpAt, Q performs selective attention, causing the computation to drop sharply with sparse ratio. However, different Q independently select distinct KV blocks, and all of these KV blocks together also incur high memory access. 
Fig.~\ref{fig:challenge}(b) illustrates these dispersed KV selections in light blue, along with the nearly unchanged KV memory access volume with full attention. But chiplet technology cannot enhance memory bandwidth, calling for a revision of the distributed attention mechanism for SpAt to exploit its NoI advantage.


For decoding of each turn, CP causes a significant inter-rank memory access imbalance in SpAt, as shown in Fig.~\ref{fig:challenge}(c). We collect the data from Llama-3.1 8B under an open source SpAt implementation \cite{fsa} at 0.95 sparsity with a 128k cache length, and CP=8. The memory-access volumes of the ranks are sorted and then averaged across tokens, layers, and heads. It is observed that memory access imbalance is significant and widespread, limiting the effective bandwidth utilization of all memory chiplets.  
Such a severe imbalance must arise from inherent cause, rather than sparsity randomness. Therefore, the imbalance must be fundamentally addressed through optimization.

%% file: 4_design.tex
\section{Optimization for Partial-Prefill}
\label{sec:adaption}

\begin{figure}[t]
  \centering
  \includegraphics[width=\linewidth]{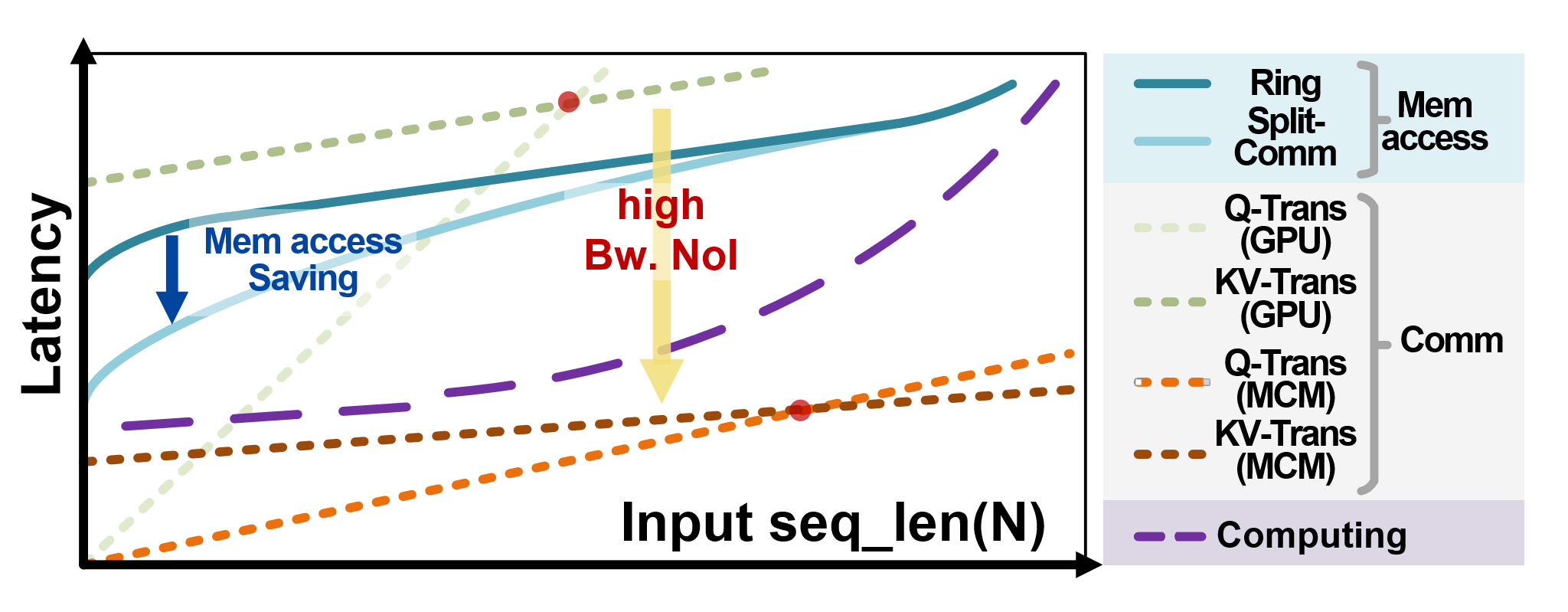}
  \vspace{-15 pt}
  \caption{Comm, Comp, and Memory Access latency trends. 
  }
  \label{fig:adaptive}
  \vspace{-10 pt}
\end{figure}

%
\subsection{Rethink of Ring Attention}
For partial-prefill, the basic Ring Attention needs modification to accommodate MCM architectures and diverse inference scenarios.
To extend the basic Ring Attention for multi-turn inference, we start with the general case where the cached context length is $P$ and the new input sequence length is $N$.
For communication, Ring Attention can be executed in two ways: \textbf{exchanging either Q or KV across dies via the ring is sufficient to complete attention for all tokens}. In the basic Ring Attention with no context cache ($P=0$), KV-transfer is used to reduce communication volume since KV has fewer heads than Q in GQA LLMs. 
However, as the amount of Q only depends on new input tokens ($N$), the relative sizes of KV and Q are no longer fixed when $P$ is large. Q-transfer has lower volume if the following holds:
\begin{equation}
\label{Eq:ring_q}
\frac{N}{P}<\frac{2h_{KV}}{h_{Q}}  
\end{equation}

Based on the above minor modification of the implementation, the bottleneck of attention shifts with the input sequence length $N$. Fig.~\ref{fig:adaptive} illustrates how the latencies of communication, memory access, and computation scale with $N$ under a fixed cache length $P$. The dashed lines represent the communication latency of Q-transfer and KV-transfer. Q-transfer has a steeper slope with $N$ for more heads, while KV-transfer has a larger intercept due to long cache. Their intersection corresponds to the critical point of Eq.~\ref{Eq:ring_q}, and different transfer strategies can be selected on either side of it. 

In most multi-turn long-context partial-prefill cases, communication is not the efficiency bottleneck. In agent scenarios, the average input sequence length per turn is reported to be 429 \cite{dualpath}; therefore, we need to transfer only the \textit{Q} of these few tokens. However, these tokens still require attention against the cached context spanning hundreds of K tokens, leading to heavy memory access volume even under SpAt. As shown in Fig.~\ref{fig:adaptive}, communication latency is lower than that of memory access for such moderate $N$ and large $P$. In particular, for the MCM, the communication curve further shifts downward compared to conventional GPU clusters; this is attributed to more efficient NoI, which further mitigates communication overhead. Cutting-edge interconnects for large-scale GPU clusters, like NVL72 \cite{nvidia2024blackwell}, produce analogous outcomes.

These emerging hardware and software features undermine the main advantage of Ring Attention, i.e., communication can be overlapped with computation and memory access to hide its latency. As described in Sec.~\ref{sec:challenge}, SpAt reduces the arithmetic intensity of attention, rendering memory access a more prominent bottleneck. However, \textbf{the communication-friendly Ring Attention may incur redundant memory access,} especially for a modest $N$. Ring Attention divides the attention of all Q into CP timesteps, causing repeated KV accesses across timesteps. As shown in Fig.~\ref{fig:a2a}(a), $Q_{0}$ and $Q_{1}$ arrive at rank 0 at timesteps 1 and 2. If they both select the same KV block (block 4 in Fig.~\ref{fig:a2a}(a)), it must be read twice over two timesteps. Therefore, Ring Attention may not be well-suited for multi-turn inference on the MCM architecture.

\subsection{Bi-mode Distributed Sparse Attention}
\label{sec:spli-comm Attention}
\begin{figure}[t]
  \centering
  \includegraphics[width=\linewidth]{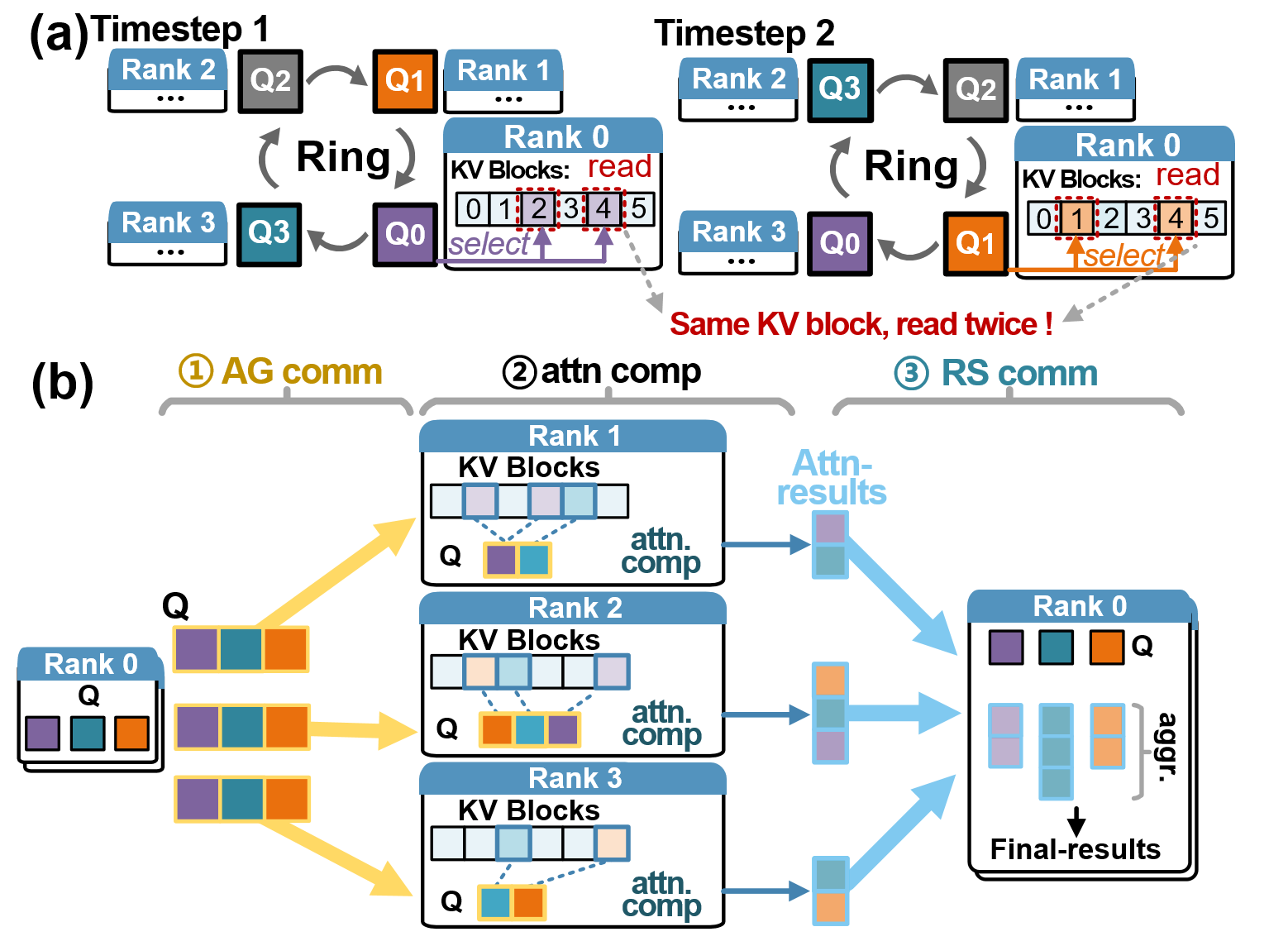}
  \vspace{-10 pt}
  \caption{Split-Comm distributed attention. 
  }
  \label{fig:a2a}
  \vspace{-15 pt}
\end{figure}
To maximize efficiency, each fetched KV block should be attended by as many Q as possible. Therefore, partitioning of input tokens across timesteps should be avoided.
To this end, we separate communication from attention computation, termed \textit{Split-Comm Attention}. As shown in Fig.~\ref{fig:a2a}(b), we send Q to the ranks hosting their selected KV blocks at first. In the worst case, each Q must be sent to all other ranks, corresponding to All-Gather (AG) communication. Then, ranks compute the attention of their hosting KV blocks. All Qs that select a given KV block are merged for joint computation. Such KV-block–centric SpAt is already supported by the kernel in \cite{fsa}. Finally, the partial attention results of each Q are aggregated by Reduce-Scatter (RS) communication. \textbf{Split-Comm minimizes redundant kv cache memory access, but the communication latency cannot be overlapped.} However, this leverages the high-bandwidth, large-scale NoI of MCMs, yielding overall benefits when communication latency is much shorter than attention.


Split-Comm does not always outperform Ring Attention. With a large $N$, the numerous Q form a massive tensor that requires deep tiling to fit within the on-chip SRAM. Tiling and the timesteps in Ring Attention both slice the computation, incurring similar redundant memory accesses. 
Fig.~\ref{fig:adaptive} also shows the memory access of Split-Comm: it is lower than Ring when $N$ is small, but gradually approaches the Ring as $N$ increases. When $N$ is large, the communication overhead of Split-Comm leads to negative gains. 

Therefore, Split-Comm is typically paired with Q-transfer when $N$ is not excessively large. In such cases, adopting Ring Attention necessitates the transfer of both $Q$ and the partial attention results, incurring a total communication volume of: $2Nd_hh_q*(CP-1)/CP$. Split-Comm decouples the same communication into two stages, maintaining a total traffic volume consistent with Ring Attention. Therefore, Split-Comm does not introduce additional communication volume overhead. Its cost just lies in the non-overlapping nature of communication relative to attention. Thus, split-comm does not incur significant communication latency in appropriate scenarios while improving total efficiency by reducing KV cache access, as detailed in Sec.\ref{sec:evaluation}.

We adopt Ring Attention and Split-Comm as a Bi-Mode method, using length-driven adaptation to select the attention mechanism under varying $N$ and $P$, as described in Alg.~\ref{alg:ada}. 
Here $th$ is a critical threshold and when $N<th$ it satisfies:
\begin{equation}
\label{Eq:th}
L_{c}+\frac{Mem(Q,\frac{KV}{CP})}{BW_{m}}<max(L_{c},\frac{CP\cdot Mem(\frac{Q}{CP},\frac{KV}{CP})}{BW_{m}} ) 
\end{equation}

The Q, KV, $L_{c}$ in Eq.~\ref{Eq:th} are functions of $N$ and $P$. The latencies of Ring and Split-Comm attention are estimated on the two sides of the Eq.~\ref{Eq:th}. $L_{c}$ denotes the communication latency, which increases linearly with $N$. The function $Mem$ denotes the memory-access volume in the attention of given $Q$ and $KV$, $BW_{m}$ denotes the memory bandwidth. For Ring Attention, $Q$ and $KV$ are divided by $CP$, representing $CP$ timesteps. And the \textit{max} operation reflects the overlap in Ring Attention. $Mem$ function is influenced by tiling strategies, which can be determined by a mapping optimizer \cite{timeloop}. For simplicity, Eq.~\ref{Eq:th} omits computation latency concurrent with memory access. Although Eq.~\ref{Eq:th} is complex, the threshold $th$ can be determined via offline testing under fixed model and hardware configurations. 
Our experiments indicate that $th$ generally exceeds the mean input length of agent application scenarios \cite{dualpath}, demonstrating that split-comm holds a performance advantage over Ring Attention in the majority of cases. Detailed evaluations are in Sec.~\ref{sec:evaluation}.





\begin{algorithm}[t]
\caption{Length-driven SpAt Adaptation}
\label{alg:ada}
\SetAlgoNlRelativeSize{0}
\SetAlgoNlRelativeSize{-1}


    \If{$N \geq th$}{
        \eIf{$\frac{N}{P}<\frac{2h_{KV}}{h_{Q}}$}{
            Q-transfer Ring Attention
        }{
            KV-transfer Ring Attention
        }
    }{
        \textbf{else} Split-Comm Attention
    }

\end{algorithm}
\section{Optimization for Decoding}
\subsection{Interleaved Token Assignment}
\label{sec:imbalance}

As mentioned in Sec.~\ref{sec:challenge}, SpAt under CP introduces severe KV-cache access imbalance during decoding. This imbalance degrades performance by preventing full utilization of memory chiplets. 
\textbf{The cause of imbalance lies in the conflict between KV selection probability spread in SpAt and the token assignment in distributed attention}. 

\begin{figure}[t]
  \centering
  \includegraphics[width=\linewidth]{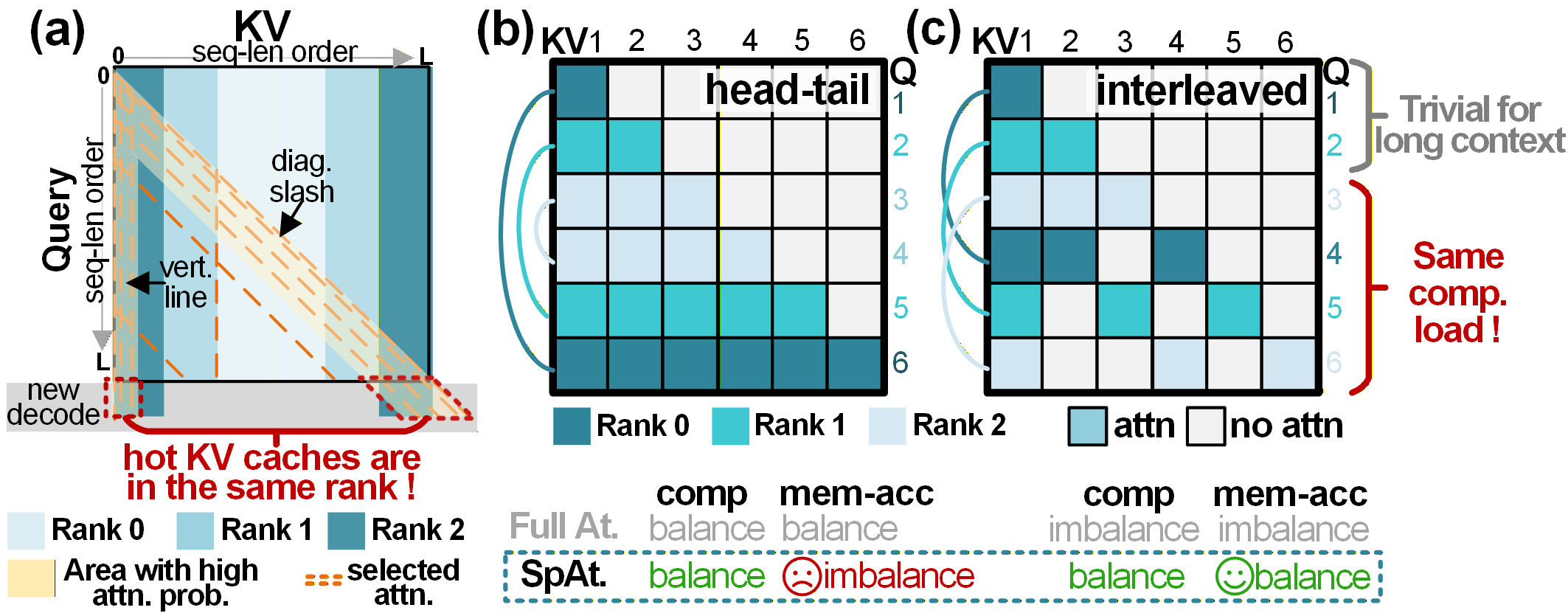}
  \vspace{-15 pt}
  \caption{Imbalanced memory access in the decoding stage.
  }
  \label{fig:imbalance}
  \vspace{-15 pt}
\end{figure}

Fig.~\ref{fig:imbalance}(a) illustrates the spread of SpAt in a long context, with most selected attention shown as orange dashed lines and blank areas indicating no attention between corresponding Q and KV. 
Previous research \cite{minference} reveals that the SpAt pattern of LLMs exhibits a $\wedge$-shape structure, as shown by the yellow regions in Fig.~\ref{fig:imbalance}(a). The Q–KV token pairs in these regions have higher attention probabilities, covering most attention selections. The $\wedge$-shape consists of two parts. The vertical line indicates that all tokens tend to attend to the beginning of the sequence, i.e., \textit{token sink} \cite{streamingllm} phenomenon; the diagonal slash region represents tokens preferring to attend to their local neighboring context. 
For decoding, the new token appears at the sequence end, shown as an additional gray row at the bottom of Fig.~\ref{fig:imbalance}(a). Based on the $\wedge$-shape, this token is most likely to attend to the KV at the beginning and end of the context, circled in red as the hot KV cache.

The above property of SpAt causes hot KV caches to be unevenly distributed across memory chiplets in terms of selection probabilities, further worsened by the currently used token assignment. KV cache in decoding is generated during prefill; thus, the token assignment in prefill determines which KV caches are grouped on the same rank.
As mentioned in Sec.~\ref{sec:para}, common distributed attention employs head-tail token assignment for workload balancing. 
Therefore, the KV cache at the beginning and end of the context is assigned to the same CP rank, which precisely has the highest selection probability. Fig.~\ref{fig:imbalance}(a) uses different background colors to indicate which parts of the KV cache are stored on each ranks. All hot KV caches are located in rank 0. Other ranks store KV cache closer to the sequence middle, where selection probability gradually decreases. Thus, KV caches on each rank naturally have uneven selection probabilities during decoding, leading to varying amounts of KV caches being accessed.

A fundamental solution to the memory access imbalance demands a change in token assignment. 
Fig.~\ref{fig:imbalance}(b) shows the head-tail assignment for compute workload balancing under full attention. 
We note a key insight: \textbf{In SpAt, each token attends to top-k KV blocks with a fixed count, so the computation does not vary with token position}. As shown in Fig.~\ref{fig:imbalance}(c), Q3–Q6 each attend to three KV and therefore have the same computation workload. Thus, we do not need to follow head-tail token assignment. We employ the solution with \textit{interleaved token assignment} in Fig.~\ref{fig:imbalance}(c), allocating tokens to CP ranks alternately. Since each rank is assigned tokens spanning the full sequence, the KV caches on each rank have equal total selection probabilities in subsequent decoding. Thereby, the KV memory access imbalance is mitigated in decoding. Moreover, due to the property of SpAt, interleaved token assignment preserves balanced prefill computation, which is not achievable under full attention. Although the beginning tokens incur less computation due to the too short preceding context, their proportion is negligible in a long context. In practice, interleaving is performed at the granularity of KV block length $B_{k}$. 

\subsection{Agile Cache Balancing}
\begin{figure}[t]
  \centering
  \includegraphics[width=\linewidth]{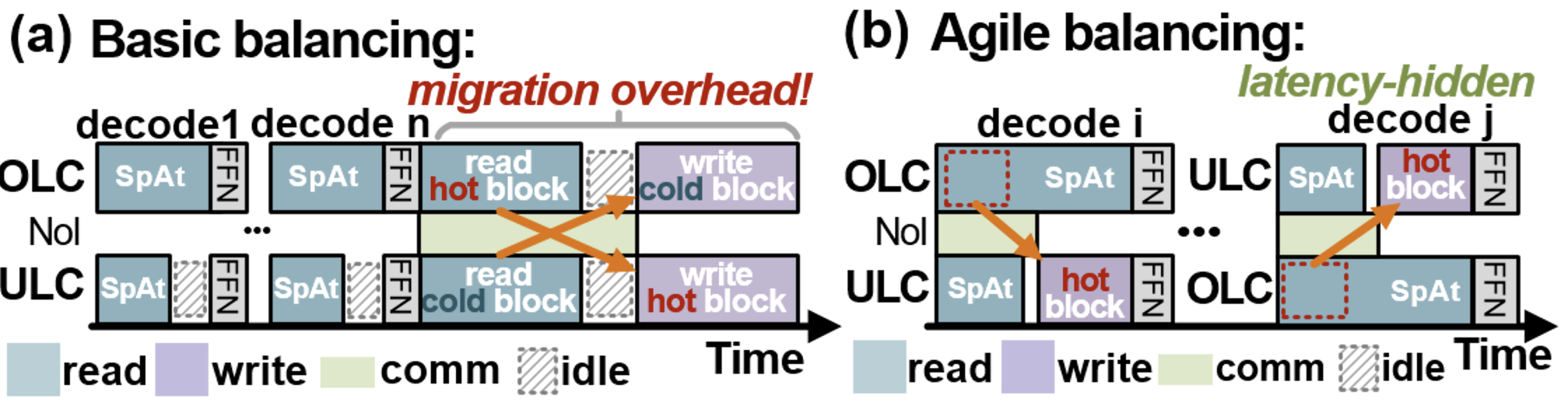}
  \vspace{-15 pt}
  \caption{Dynamic KV cache balancing methods. 
  }
  \label{fig:agile}
  \vspace{-15 pt}
\end{figure}
While static interleaved token assignment addresses the inherent cause of KV access imbalance, it fails to account for the access skews induced by shifting SpAt focus.
Our experiments show that the cache access for the peak-load chiplet remains 52.8\% higher than the ideal balanced cases. To further enhance memory access efficiency, dynamic balancing is required for online KV cache redistribution. However, naive redistribution may introduce massive KV blocks exchange, incurring additional communication and read/write overhead. To minimize the costs, we propose \textit{agile cache balancing} for hot KV block selection and latency-hidden migration. 

Selection is required to migrate only the hot blocks most likely to be accessed from an overloaded chiplet (OLC) to the underloaded chiplet (ULC). A key property is leveraged: KV block selection exhibits significant similarity across adjacent decoding iterations. 
However, tracking and sorting access frequencies for tens of thousands of blocks is complex. We adopt the Space-Saving \cite{space_saving} algorithm to select the hot blocks. A hot block table (HBT) is maintained, whose entries are much fewer than the total count of KV blocks in a chiplet. Each entry stores a target KV block index and its access count. During each decoding iteration, the HBT is updated for each activated KV block based on the rules: (1) If the block index hits an existing entry, its counter is added by one. (2) If the index is not in HBT and HBT is not full, this index is inserted into HBT with its counter set to one. (3) If the index misses and HBT is full, the index with the minimum counter is replaced by the new block index, and the counter is increased by one from its previous value. After a few iterations (e.g., 50), all counters in HBT are divided by 2 to fade out the influence of older records. This algorithm reduces the number of blocks to be tracked and has been proven effective in prior scenarios \cite{row_hammer}. In our implementation, the number of HBT entries is set to 22.5\% of the total blocks stored per chiplet.

Fig.~\ref{fig:agile}(a) shows the basic balancing approach, which triggers a moderate-sized KV block migration at fixed intervals of decoding iterations: an OLC sends hot blocks with data volume $D_m$ to a ULC, which reciprocates cold blocks also with $D_m$ volume back to OLC. The basic balancing suffers from two drawbacks. First, notable migration overhead is introduced. Second, balancing at fixed intervals often lags behind hot KV variations. The SpAt focus shifts unpredictably, making it difficult to determine an ideal interval. We identify three optimization opportunities. First, the top blocks in HBT have high probabilities of being activated in SpAt. By limiting migration to blocks already fetched in SpAt, the additional read volume of the OLC can be avoided. Second, reciprocation—intended to keep uniform memory occupancy—is unnecessary. By limiting $D_m$ to a minimal size, the influence on memory occupancy can be negligible. Moreover, in long-term inference, the shifting roles of chiplets as OLC or ULC ensure that migration-induced occupancy variation does not accumulate over time. Third, ULC has idle time in SpAt without any memory read/write, which is under-utilized. 

To minimize migration overhead, we implement the latency-hidden migration as shown in Fig.~\ref{fig:agile}(b). \textbf{Migration is moved from the intervals between iterations and fused with SpAt within each iteration.} Upon fetching the activated blocks during SpAt, the OLC filters blocks with total volume $D_m$ based on the HBT counter for migration, and releases the memory space of the migrated blocks. After receiving the blocks, ULC buffers them on-chip without any reciprocation to OLC. \textbf{Since the ULC completes SpAt earlier due to fewer memory accesses, its idling provides a time window to write hot blocks from buffer to memory.} Migration latency is hidden within the SpAt provided that:
\begin{equation}
\label{Eq:agile1}
L_c + \frac{D_{m}}{BW_c} + \frac{D_{m}}{BW_m} < T_{O}
\end{equation}
\begin{equation}
\label{Eq:agile2}
T_{U} + \frac{D_{m}}{BW_m} < T_{O}
\end{equation}
where $T_{o}$ and $T_{u}$ are the SpAt latency of OLC and ULC, $BW_c$ and $BW_m$ denote the communication and write bandwidths, and $L_c$ is the fixed communication latency. Owing to the low latency of NoI, the single-hop end-to-end $L_c$ can be less than 75 ns \cite{NoI_lc}. Compared with $\mu$m-scale SpAt latency, $L_c$ does not noticeably congest the available time window, rendering latency-hiding achievable when $D_m$ is low.
Since the memory access loads are deterministic before SpAt, such precise orchestration is feasible. Although per-migration $D_{m}$ is minimal, the near-zero overhead enables balancing to be performed at every iteration. This feature also addresses the second drawback of basic balancing, enabling instant reaction to shifting SpAt focus, rather than waiting for fixed long intervals. Essentially, our approach partitions the basic balancing into minimal migrations and exploits the overlooked SpAt idle window for latency-hiding.

In implementation, a chiplet is classified as OLC if the ratio of actual KV access volume to the expected-mean KV access volume exceeds a threshold (e.g., 1.25). Non-OLC chiplets are all treated as ULCs. We restrict that OLC can only migrate blocks to a neighboring ULC in the same CP group. Due to rapid balancing, adjacent chiplets do not all persist as OLCs across many iterations. Therefore, our policy does not lead to long-term migration stalls.

%% file: 5_evaluation.tex
\section{Evaluation}
\label{sec:evaluation}
\subsection{Evaluation Setup}
We evaluate the performance using the distributed system simulator ASTRA-sim \cite{astra-sim}, which integrates the NS3 \cite{ns3} network simulator. We also use Ramulator 2.0 \cite{ramulator2} to model memory access, and Timeloop \cite{timeloop} for tiling optimization. The SpAt selection is based on runtime traces.

Our default MCM configuration is based on \cite{WSC_LLM}: the MCM consists of an 8 $\times$ 8 die array and employs a parallel strategy of HP=8 and CP=8. Each die contains a 20mm $\times$ 20mm logic chiplet with 600TFLOPS (FP8) and 20MB on-chip global buffer, and two HBM3 \cite{hbm3} chiplets providing 2TB/s bandwidth. Double-buffering is utilized to overlap computation with memory access latency. Under this configuration, each die is approximately half the scale of an advanced cloud GPU, while the reduced chiplet area enhances manufacturing yield. The chiplet die-to-die bandwidth density and latency are based on \cite{dojo,NoI_lc}. According to the size of a logic chiplet and subtracting the length occupied by HBM3, each logic chiplet edge provides 1TB/s bidirectional NoI bandwidth. Given the memory capacity constraints of HBM during long-context inference, a batch size of 8 is employed.

Each GPU-based configuration has 32 GPUs. Each GPU provides twice the compute and memory resources of an MCM die, ensuring that the total resources match those of the MCM. GPU-basic utilizes 8-GPU NVLink nodes \cite{nvidia2022hopper} and IB \cite{deepseekai2025deepseekv3technicalreport} for inter-node connectivity. GPU-NVL72 employs the latest NVL72 \cite{nvidia2024blackwell} technology to connect all 32 GPUs with 1.8TB/s bandwidth. For a fair comparison, GPU-NVL72 uses the same per-device configurations as GPU-basic rather than the official NVL72 specs. Llama 3.1 405B \cite{llama3} is selected as the target model with the SpAt implementation from \cite{fsa}. 
The KV block length ($B_{k}$) equals 64 \cite{fsa}, the sparse ratio equals 0.85.

\subsection{Evaluation on Context Parallelism}

\begin{figure}[t]
  \centering
  \includegraphics[width=\linewidth]{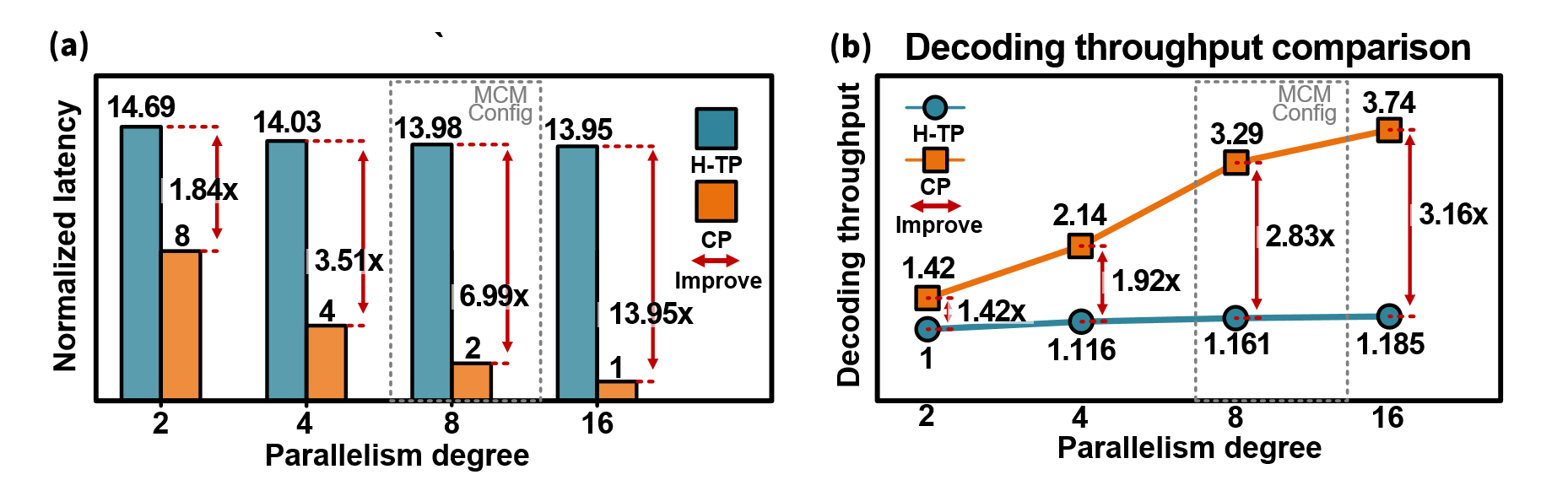}
  \vspace{-15 pt}
  \caption{Performance gains of CP over H-TP.  
  }
  \label{fig:eva_cp}
  \vspace{-5 pt}
\end{figure}

To demonstrate the advantage of CP over H-TP, we first evaluate the latency of random KV cache access under two parallelisms. 
We temporarily assume that the KV cache access volume is equal across memory chiplets under CP to avoid imbalance being a confounding factor.
As shown in Fig.~\ref{fig:eva_cp}(a), the x-axis denotes the parallel degree of CP or H-TP, and the y-axis denotes the normalized memory access latency. It can be observed that CP achieves substantially higher memory-access efficiency than H-TP across all groups. Under the MCM configuration, H-TP’s read latency reaches 7$\times$ that of CP.
This reveals the harmful impact of the memory fragmentation caused by H-TP. We even observe that once H-TP reaches 4, further increasing H-TP (i.e., using more chiplets) yields little improvement in KV cache access. This implies that the increased memory fragmentation from higher H-TP negates the total bandwidth gains from more memory chiplets, severely limiting H-TP’s scalability. 


To demonstrate the impact on performance, we focus on the decoding phase that is strongly tied to KV cache access, and evaluate the throughput of H-TP and CP under different parallelism degrees. Memory access imbalance of CP (with our optimization in Sec.~\ref{sec:imbalance}) is taken into account, and the results are shown in Fig.~\ref{fig:eva_cp}(b). For the default MCM configuration with CP=H-TP=8, the decoding throughput of H-TP is only 35.7\% of that of CP, which is unacceptable. As the degree of parallelism increases, 
CP maintains nearly linear gains in decoding throughput, and the throughput gap is widened. The above results reveal that CP is essential for large-scale MCMs, underscoring the value of our optimizations. 

\subsection{Evaluation on Partial-Prefill Optimization}
\label{sec:eva_len}

\subsubsection{Memory Access Saving of Split-Comm} 
\begin{figure}[t]
  \centering
  \includegraphics[width=\linewidth]{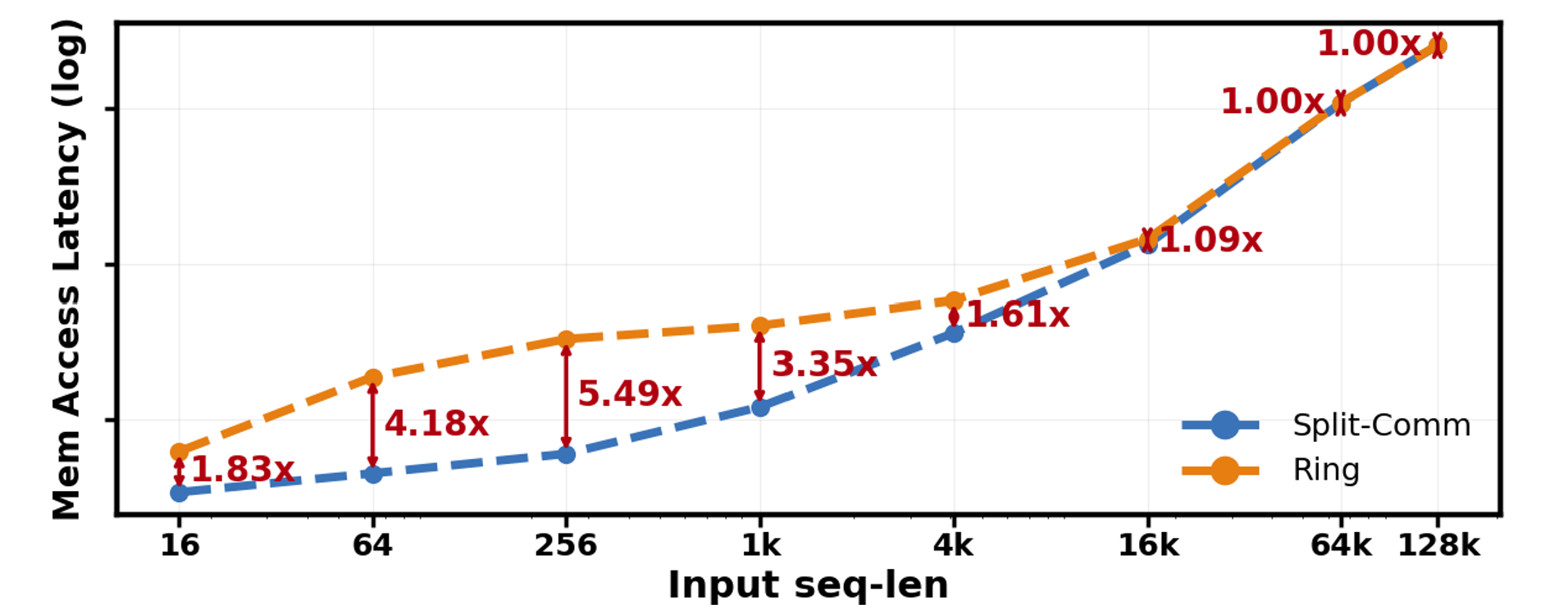}
  \vspace{-15 pt}
  \caption{Memory access saving of Split-Comm.}
  \label{fig:eva_mem}
  \vspace{-10 pt}
\end{figure}

Fig.~\ref{fig:eva_mem} compares the KV cache access latency of Ring Attention and Split-Comm for $P$=512k as $N$ increases from 16 to 128K. It can be seen that in SpAt, Split-Comm reduces KV cache memory access by up to a notable 5.49$\times$ when $N=256$, which is close to the typical partial-prefill length in applications \cite{dualpath}. But the memory access saving varies with $N$ in a non-trivial trend. 
When $N$=16, such a small number of Q select different KV blocks without overlap, which limits the benefit of token merging in Split-Comm attention. As $N$ increases, overlapped KV selections become prevalent, causing Ring to fall far behind Split-Comm in the middle of Fig.~\ref{fig:eva_mem}. Within this range, Ring Attention requires a near-complete traversal of KV cache during every timestep, whereas Split-Comm facilitates the reuse of the KV cache across more Q. As $N$ exceeds 4k and continues to scale, the memory latency gap between the two methods gradually diminishes. In such cases, both Q and KV are large, requiring tiling to fit within the on-chip SRAM. Tiling also partitions attention between tokens into steps, causing the practical implementations of Split-Comm and Ring Attention to gradually converge.

\begin{figure}[t]
  \centering
  \includegraphics[width=\linewidth]{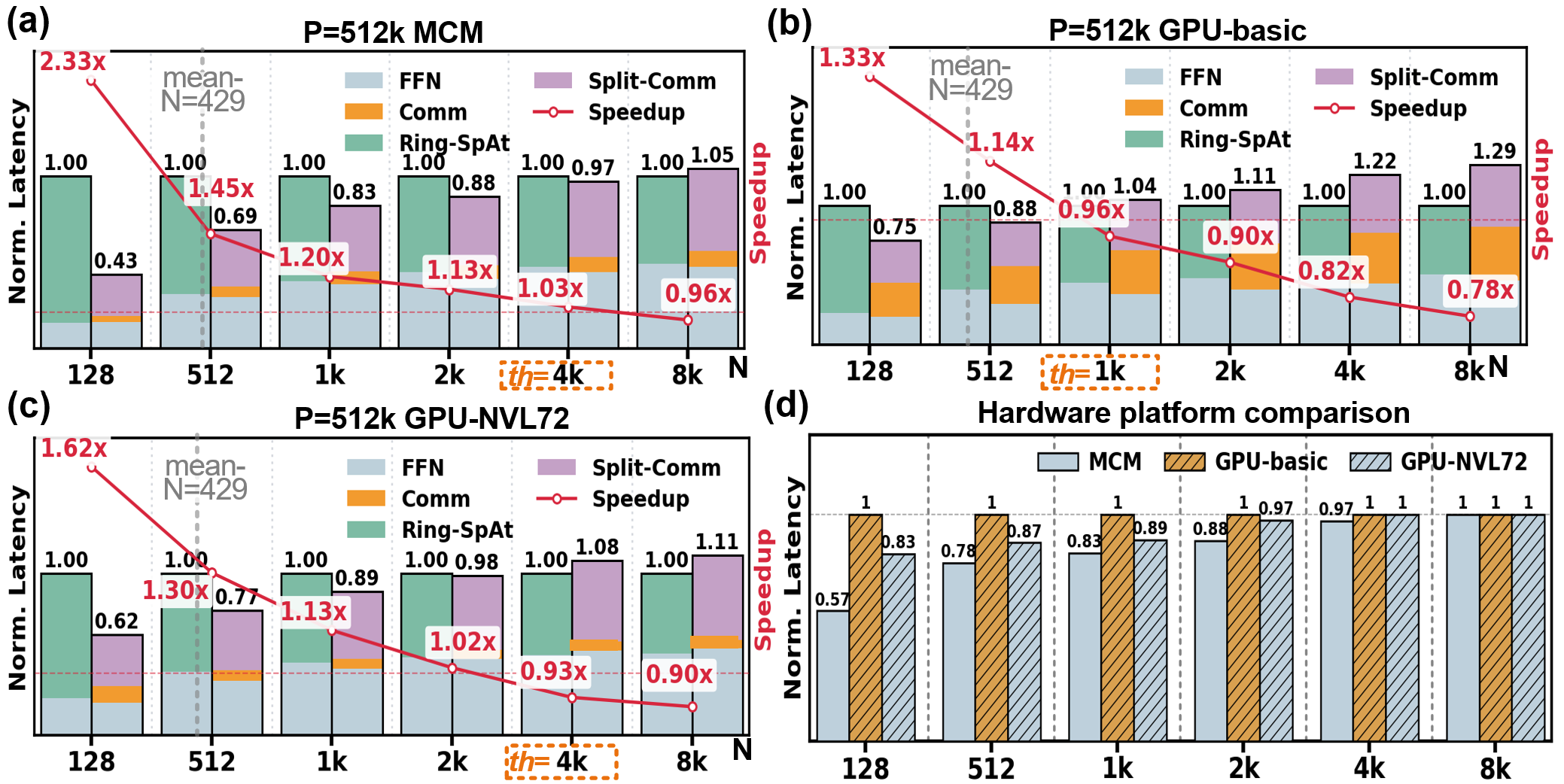}
  
  \vspace{-10 pt}
  \caption{Gains of Bi-Mode SpAt for partial-prefill.}
  \label{fig:eva_prefill}
  \vspace{-10 pt}
\end{figure}

\subsubsection{End-to-End Partial-Prefill Latency}
We evaluate partial-prefill latency on different hardware platforms with a long KV cache ($P$=512k) and new inputs of varying length $N$, as shown in Fig.~\ref{fig:eva_prefill}(a)-(c). For each $N$, the left bar represents the total latency using Ring Attention (normalized to 1), while the right bar denotes the relative total latency using Split-Comm. Colored segments within each bar indicate the latency breakdown. The red line represents the speedup ratio achieved by Split-Comm.
 As shown in Fig.~\ref{fig:eva_prefill}(a), Split-comm significantly enhances MCM under moderate-$N$ regimes. When $N=512$—which is near the average partial-prefill length in practical applications—it achieves a 1.45$\times$ speedup. As $N$ increases, the speedup diminishes due to reduced memory savings, increased communication overhead, and FFN layer dilution. Split-Comm even yields negative returns beyond the threshold $th$ in Eq.~\ref{Eq:th}.
 Due to the NoI advantages, the $th$ can be maintained up to $N$=4k. For GPU-basic in Fig.~\ref{fig:eva_prefill}(b), Split-Comm achieves a 1.33$\times$ speedup at $N=128$, but communication overhead shifts the $th$ to about $N=1k$. Nevertheless, the benefit range also covers considerable application scenarios, showing the generalizability of Split-Comm on GPUs. For GPU-NVL72, split-comm exhibits a broad range of benefits like MCM as shown in Fig.~\ref{fig:eva_prefill}(c). A minor difference lies in a slight decrease in speedup for small $N$, due to the higher fixed communication latency of NVL72. This result demonstrates that Split-Comm yields promising gains in next-generation GPU clusters with enhanced interconnections. 

Ring Attention outperforms Split-Comm with short context and long input sequence. Specifically, at $N=64\text{k},P=4\text{k}$, the Split-Comm suffers a 35\% efficiency loss even on the MCM. Such cases are rare in multi-turn inference, yet they underscore the need for a Bi-mode method to adapt to varying sequence lengths.

Fig.~\ref{fig:eva_prefill}(d) compares the latency of MCM, GPU-basic, and GPU-NVL72 using Bi-mode attention. By leveraging communication advantages, MCM and NVL72 significantly outperform GPU-basic for $N < 2\text{k}$. For larger $N$, the latencies are compute-bound and converge as three configurations share the same compute performance.


\subsection{Evaluation on Decoding Optimization}
\begin{figure}[t]
  \centering
  \includegraphics[width=\linewidth]{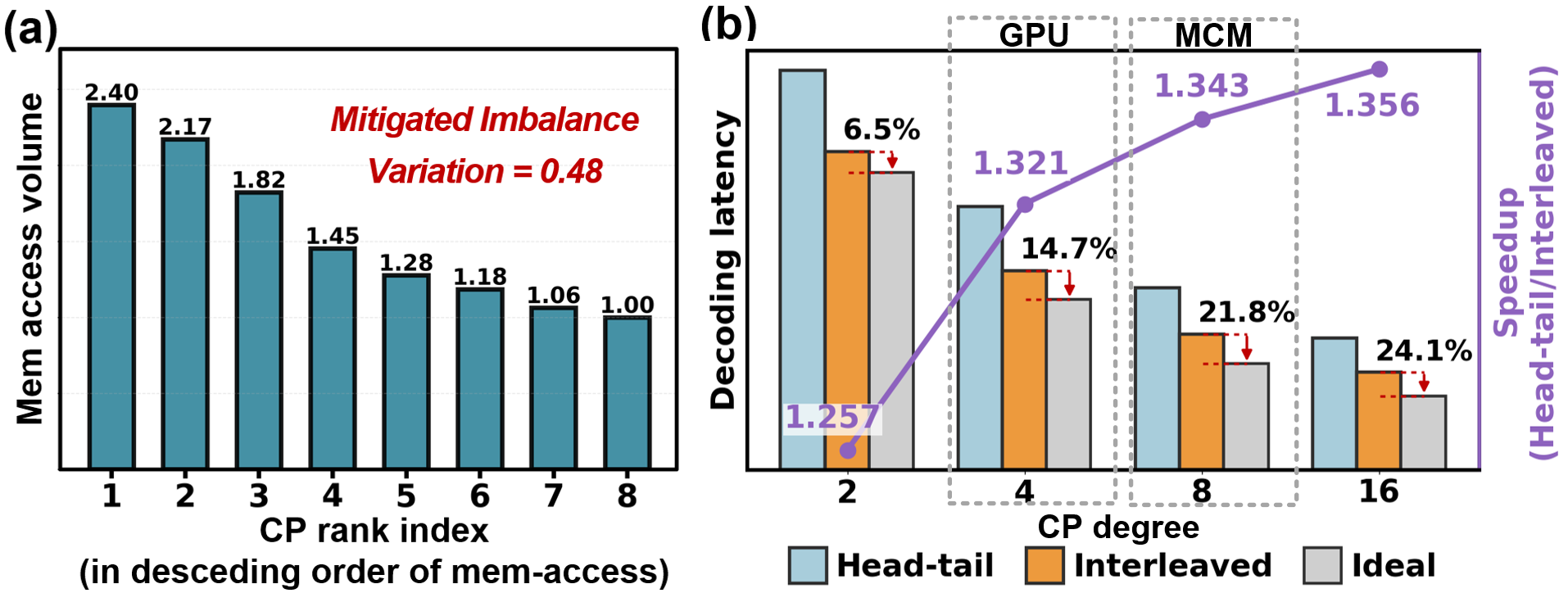}
  \vspace{-15 pt}
  \caption{Balancing effect of interleaved token assignment.}
  \label{fig:eva_decode}
  \vspace{-15 pt}
\end{figure}
\subsubsection{Evaluation for Interleaved Token Assignment}
To evaluate the effect of interleaved token assignment, we evaluated the average memory access disparity across CP ranks in Fig.~\ref{fig:eva_decode}(a), using the same setup and method as in Fig.~\ref{fig:challenge}(c). 
Compared to the head-tail assignment, interleaved assignment significantly reduces the average KV access variance from 2.775 to 0.48. 
Fig.~\ref{fig:eva_decode}(b) illustrates the end-to-end decoding performance improvement at $P=512K$. Despite the default configuration with CP=8, we assess other scales to show the benefits at varying CP degrees.
It can be observed that interleaved token assignment achieves a significant 1.34$\times$ speedup for MCM. The GPU groups also achieve 1.32$\times$ speedup, showing the generality of our optimization. The improvement grows with increasing CP, as each memory chiplet is assigned fewer KV-cache blocks, making KV block-level variations more directly reflected in overall memory access. Compared to the ideal balanced case, the static method still has 21.8\% room for improvement. 

\subsubsection{Evaluation for Agile Token Balancing}
\begin{figure}[t]
  \centering
  \includegraphics[width=\linewidth]{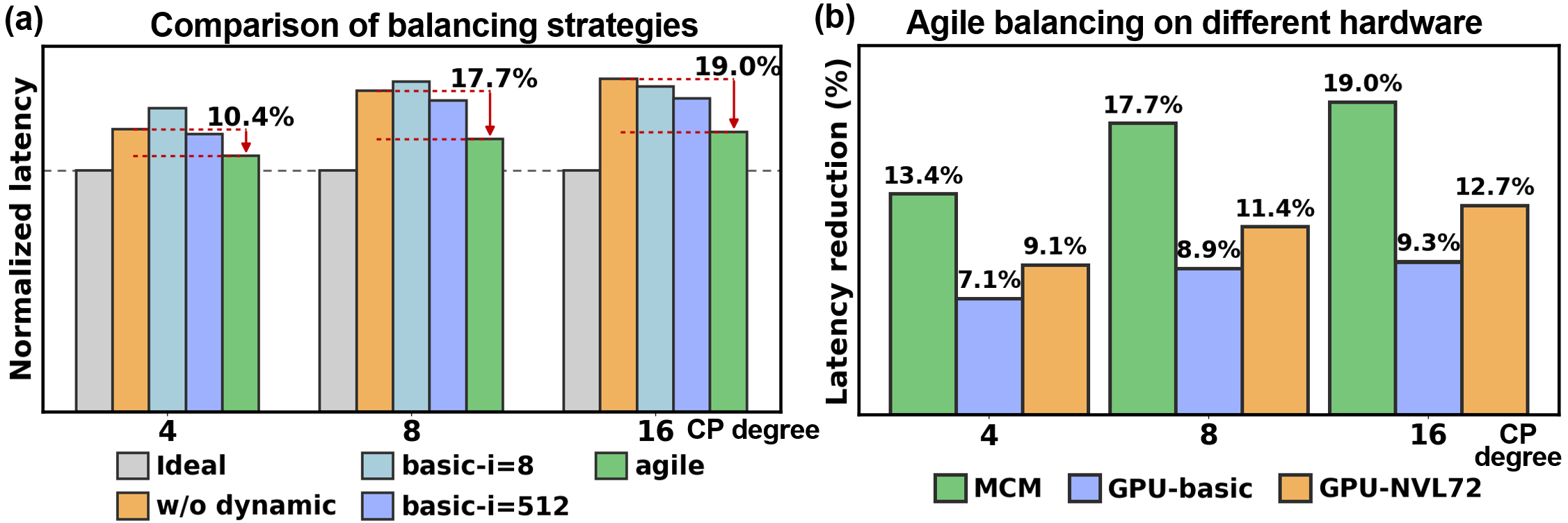}
  \vspace{-15 pt}
  \caption{Enhancement of agile cache balancing.}
  \label{fig:eva_agile}
  \vspace{-15 pt}
\end{figure}
Fig.~\ref{fig:eva_agile}(a) compares decoding latencies of dynamic KV-balancing on the MCM with different CP degree, based on the interleaved token assignment and $P=512$k. 
For basic balancing, OLC is assumed to migrate 10\% of KV blocks at variable intervals ($I$). Basic balancing either incurs excessive migration overhead at $I=8$ or lags behind SpAt shifts at $I=512$. By hiding migration and rapid adapting, agile token balancing outperforms all basic balancing strategies. At $CP=8$, it further improves decoding efficiency by 17.7\% compared to the baseline, nearly reaching the ideal balanced state.

We also evaluate the effectiveness of agile cache balancing across different hardware platforms. As illustrated in Fig.~\ref{fig:eva_agile}(b), the low-latency NOI provides wider migration time windows on the MCM, being well-suited for the fine-grained migrations in agile balancing. Although NVLink has higher communication latency \cite{NVlink_lc}, agile balancing still reduces latency by 8.9\% and 11.4\% on GPU-basic and GPU-NVL72 when CP=8, respectively. Interleaved token assignment and agile cache balancing together yield 1.53$\times$, 1.32$\times$, and 1.36$\times$ average decoding speedup across MCM, GPU-basic, and GPU-NVL72.
\section{Conclusion}
In this work, we explore long-context LLM inference with SpAt on large-scale MCM. We employ context parallelism to avoid fragment memory access of KV cache. For partial-prefill, we propose Split-Comm Attention, which leverages the high-bandwidth NoI of MCMs to trade communication for memory access savings. For decoding, we reveal the inherent cause of memory access imbalance and exploit SpAt’s properties to employ interleaved token assignment. Agile cache balancing is also proposed to migrate memory access imbalance dynamically.  


